# Rapid electron trapping studied by pump-probe photoconductivity: kinetic analysis.

Ilya A. Shkrob [a] and Leonid Ryzhik [b]

[a] *Chemistry Division, Argonne National Laboratory, Argonne, IL 60439*

[b] *Department of Mathematics, University of Chicago, Chicago, IL 60637*

## Abstract

The use of ultrafast pump-probe conductivity (PPC) for studies of photoelectron dynamics in nonpolar molecular liquids has been based upon the perturbation of geminate recombination dynamics of trapped electrons by their laser photoexcitation into the conduction band. Such a method is unsuitable for the studies of electron trapping dynamics of quasifree electrons *per se*. We demonstrate that the PPC method can be extended to study such dynamics, provided that the time resolution of the conductivity setup is better than the ratio $\mu_e \tau_e / \mu_s$ of the mobility-lifetime product for the quasifree electron and the mobility $\mu_s$ of the trapped electron. For some liquids (e.g., supercritical $CO_2$) this time is sufficiently long (> 100 ns) and the standard conductivity equipment can be used. Even if the time resolution cannot be increased (due to the adverse effect on the sensitivity), the trapping dynamics can be studied provided that the trapping competes with cross recombination of quasifree electrons with holes in the solvent bulk. Since the mobility of these quasifree electrons is very large (10-$10^3$ cm$^2$/Vs), this recombination is facile even when the density of ionization events is fairly low (< 1 μM).

Perturbation of the geminate electron-hole dynamics is not required for this method to work.

___________

* To whom correspondence should be addressed: *Tel* 630-252-9516, *FAX* 630-2524993, *e-mail:* shkrob@anl.gov.

## 1. Introduction

The injection of electron into a molecular liquid, by photoionization of the solvent/solute molecules or electron photodetachment from an anion or metallic cathode, results in the familiar scenario: [1,2] First, a quasifree electron, $e_{qf}^-$, is generated in the conduction band of the liquid. Then, this electron is thermalized and localized by the solvent. The nature of the resulting species depends on the nature of the liquid. In liquids whose constituent molecules have no electron affinity and no accessible 2$p$-orbitals (such as ammonia, water, alcohols, ethers, amines, and saturated hydrocarbons), the so-called solvated (cavity) electrons are generated. These species are $s$-orbital electrons localized in the interstices between the solvent molecules; the sharing of the excess electron density by these molecules is minor. In molecular liquids whose molecules have positive electron affinity, i.e., in most organic liquids, the electrons attach to one or several molecules forming monomer or multimer solvent anions (e.g., $C_6F_6$ [3] and $CS_2$ [4]).

In water, the quasifree electron localizes very rapidly, in ca. 50 fs, [5] and fully stabilizes in < 1 ps, [5-14] forming a solvated electron with the energy ca. 2 eV below the mobility edge of the conduction band. Such an outcome is typical for other hydrogen-bonded polar liquids. For other liquids, more than one kind of negative charge carrier may coexist shortly after the ionization. In saturated hydrocarbons, the electrons in shallow traps are so close in energy to the mobility edge of the liquid (ca. 100-200 meV) [2] that thermal emission from these states into the conduction band readily occurs; that is, $e_{qf}^-$ is in rapid equilibrium with these states. [2] A solvent molecule can also be a reversible electron trap: e.g., in benzene and toluene (ArH), the quasifree electron rapidly equilibrates with the molecular anion (ArH$^-$): [2,15]

$$e_{qf}^- + ArH \rightleftharpoons ArH^- \qquad (1)$$

In pressurized aromatic liquids, rxn. (1) is shifted to the right, and the resulting anion migrates ca. 10 times faster than any other ions in these liquids, by resonant charge hopping. [15] Faster-than-diffusion hopping has also been observed for solvent anions in other liquids, such as $SF_6$, [16] $C_6F_6$, [3] $CS_2$, [4] and supercritical $CO_2$. [17,18,19] This hopping would be impossible without sharing of the negative charge between several solvent molecules, i.e., the electron in these liquids rapidly samples different molecular clusters. [1] The electron involved in the equilibria similar to rxn. (1) does not have to be quasifree: In liquid acetonitrile, the dimer solvent anion $(CH_3CN)_2^-$ exists in equilibrium with the cavity electron; [20] in liquid $(CH_3)_2S$, the anion and the cavity electron coexist on the subnanosecond time scale. [21]

So far, most of the ultrafast studies have been carried out for electrons in water, [5-14,22-25] alcohols, [26] and ethers. [27,28] Pump-probe transient absorption (TA) spectroscopy was the main experimental approach. Solvation and geminate recombination dynamics of localized electrons [6-14,25] and relaxation dynamics of photoexcited solvated electrons were investigated in these liquids. [14] Molecular dynamics models have been developed [29-34] and the details of electron dynamics are gradually becoming understood. The progress in ultrafast TA studies of nonpolar liquids, such as saturated hydrocarbons, has been less spectacular, [35-41] one of the reasons being that the photoexcitation of such liquids typically yields long-lived excited states of the solvent molecules as the ionization by-product, and these states are frequently much better light absorbers than the electrons. Indeed, trapped/solvated electron in these liquids tend to absorb in the near- and mid-IR regions, a spectral region which has only recently become routinely accessible to ultrafast spectroscopy. With a few exceptions, [43,44] the research agenda was limited to the studies of recombination dynamics of fully solvated/trapped electrons. Even that limited goal was difficult to attain: The initial picosecond studies of the photoionization of saturated hydrocarbons (e.g., ref [40]) were often inadequate since the TA signal was dominated by the solvent excited states. One of the strategies for overcoming this problem was to use ultrafast mid-IR [41] and far-IR (THz) spectroscopy. [43] With the letter technique, it is possible to directly observe Drude-like quasifree electrons. Another approach (which is only suitable for thin liquid layers on metal surface) was to use ultrafast pump-probe photoelectron spectroscopy. [44] Yet another approach was to use ultrafast pump-probe conductivity (PPC). [38,39] The latter technique provides a means to selectively detect *charged* species. There is, actually, a close similarity between PPC and the more familiar 3-pulse TA spectroscopy practiced by Barbara and coworkers [14] and Schwartz and coworkers. [28] In the latter method, the change in the fraction of electrons that escape geminate recombination ("free electrons") induced by electron-detrapping laser pulse is observed by TA spectroscopy, whereas using the PPC it is observed through the dc conductivity.

## 2. PPC method for studies of geminate recombination in saturated hydrocarbons.

For geminate pairs in saturated hydrocarbons, pump-probe conductivity was first demonstrated by Braun and Scott [38] and further developed by Lukin and coworkers. [39] The solvated/trapped electron, $e_{tr}^-$, generated by UV photoionization of an aromatic solute was photoexcited into the conduction band using a short pulse of IR light:

$$e_{tr}^- \xrightarrow{h\nu_{IR}} e_{hot}^-. \qquad (2)$$

The resulting "hot" electron rapidly thermalizes, reaching the bottom of the conduction band, and localizes on the subpicosecond time scale:

$$e_{hot}^- \longrightarrow e_{qf}^- \longrightarrow e_{tr}^-. \qquad (3)$$

The rapid migration of short-lived "hot" and quasifree electrons changes the spatial distribution of trapped electrons around their parent holes. That, in turn, changes the yield of free electrons which escape the Coulomb field of their geminate partner. The IR pulse is delayed relative to the ionization UV pulse, and the increase in the relative conductivity signal from the free electrons is plotted against this delay time $\tau$.

Let $\Lambda^2$ be the mean square path for the electrons generated by photoexcitation (2) and $\mu_e^{hot} \tau_e^{hot}$ be their mobility-lifetime product. Assuming that $\Lambda$ is much less than the electron-hole separation at the delay time $\tau$ of the IR pulse, it can be shown, [45] that the relative change $\Delta\sigma_i(\tau)/\sigma_i$ in the conductivity signal $\sigma_i$ from the free (trapped) electrons

$$\Delta\sigma_i(\tau)/\sigma_i \approx \frac{\phi\Lambda^2}{6r_c^2}(1-\xi)\,\Omega(\tau), \tag{4}$$

where $\phi$ is the photoconversion, the parameter $\xi = (6k_B T/e\Lambda^2)\mu_e^{hot}\tau_e^{hot}$, and the function $\Omega(\tau)$ is given by

$$\Omega(\tau) = \int_0^\infty dr\ 4\pi r^2 p(r;\tau)\left(\frac{r_c}{r}\right)^4 \exp\left(-\frac{r_c}{r}\right), \tag{5}$$

where $4\pi r^2 p(r;t)$ is the time-dependent probability to find (trapped) electron at a distance $r$ from the parent hole at the delay time $t = \tau$, and

$$r_c = e^2/4\pi\varepsilon\varepsilon_0 k_B T, \tag{6}$$

is the Onsager radius of Coulomb interaction, where $\varepsilon$ is the dielectric constant of the solvent, $\varepsilon_0$ is the permittivity of vacuum, and $e$ is the electron charge. As seen from eq. (5), the kinetics $\Omega(\tau)$ is not directly related to the survival probability of the geminate pair. For low-mobility hydrocarbons ($\mu_e < 10^{-2}$ cm/Vs), the second term in the brackets in eq. (4) is 0.35-0.5, [38] and the net conductivity *increases* after the IR photoexcitation of the trapped electron.

The PPC method considered above is possible in two extreme situations: (i) when the electron is trapped extremely rapidly and irreversibly (as in water and alcohols) [5-16] and (ii) when the solvated/trapped electron is in a rapid equilibrium with the quasifree electron (as in saturated hydrocarbons). As mentioned in section 1, in some molecular liquids quasifree electrons undergo rapid trapping, but the time scale of this trapping can be *slower* than the geminate recombination dynamics of this electron. Below we analyze the PPC experiment in such a situation. The experimental realization of this technique on the picosecond time scale is considered elsewhere; on the slower time scale (for conversion of the solvent anion to a solute anion), a similar method was demonstrated in ref. [17]. To understand this method, it is appropriate to give a specific example.

**3. Electron dynamics in liquid-like supercritical $CO_2$.**

Ionization of liquid-like supercritical (sc-) $CO_2$ (with the critical density $\rho_c$ of 0.47 g/cm$^3$ and the critical temperature $T_c$ of 31ºC) yields two electron species: a quasifree electron, $e_{qf}^-$, and a stable multimer solvent radical anion, $(CO_2)_n^-$. [17,18,19] The quasifree electron is short-lived (the trapping time $\tau_e$ < 200 ps) and extremely mobile ($\mu_e$ > 10 cm$^2$/Vs) so that only the mobility-lifetime product $\mu_e \tau_e$ can be determined by means of dc conductivity [17] and pulse radiolysis [18] ($\mu_e \tau_e \approx 2.5 \times 10^{-9}$ cm$^2$/V at $\rho/\rho_c \approx 1.82$ and $T$=41ºC). The average thermalization path of the electron is relatively short (10-12 nm). [17,18] For $\rho/\rho_c \approx 1.8$, the solvent anion is $10^3$ times less mobile than $e_{qf}^-$ ($\mu_S$ ca. 0.016 cm$^2$/Vs); nevertheless, it has 2-10 times higher mobility than *solute* anions. [17,19] The electron photodetachment spectrum of this solvent anion corresponds to a bound-to-free transition with an onset at 1.76 eV. [17,18] The large binding energy (ca. 1.6±0.2 eV) [17] suggests that the negative charge is shared by several solvent molecules and migrates by ultrafast hopping. [1,17] The existence of a relatively long-lived quasifree electron in sc-$CO_2$ is surprising as it is known [19,46] that in low-density $CO_2$ the electron attaches to medium-size $(CO_2)_n$ clusters ($n \leq 6$) at a collisional rate. Apparently, for $\rho > \rho_c$, when the solvent conduction band emerges, the electron dynamics is quite different from that for $\rho < \rho_c$.

It is easy to see that the PPC method based on the perturbation of the geminate dynamics via the photoprocess similar to rxn. (2) cannot be used to study the electron dynamics in sc-$CO_2$. The time constant for the geminate recombination is given by the Onsager time,

$$t_c = r_c^2 / D_e, \qquad (7)$$

where $r_c$ is the Onsager radius introduced in eq. (6) and

$$D_e = (k_B T / e) \mu_e \qquad (8)$$

is the diffusion coefficient of the electron (which is much greater than that for the geminate partner). While the electron trapping time $\tau_e$ is not known, the ratio $\tau_e/t_c$ of this time and the Onsager time can be determined since this ratio (as seen from eqs. (6), (7), and (8)) is proportional to the product $\mu_e\tau_e$. For $\rho/\rho_c = 1.82$ and 41°C, $\varepsilon \approx 1.5$, $r_c \approx 35.2$ nm, and $\tau_e/t_c \approx 5.5$. Therefore, most of the electrons are trapped as anions *after* they escape the Coulomb field of the hole. Since $\mu_e \approx$ 10-100 cm²/V, [17] $\tau_e$ <5-50 ps. As demonstrated by Lukin and coworkers, [39] efficient perturbation of the geminate dynamics is possible only when the delay time $\tau$ of the excitation pulse is short with respect to the Onsager time, $\tau/t_c \approx 0.01$, i.e., the laser-induce perturbation of the geminate dynamics of quasifree electrons in sc-$CO_2$ would require femtosecond pulses. The main problem, however, is that these quasifree electrons are very poor light absorbers everywhere except for the far IR. Only *anions* (or trapped electrons) can be readily photoexcited by a short laser pulse, and these species are generated when the geminate recombination is nearly complete. Such a situation may occur in other liquids where the mobility of quasifree electron is high and its trapping time is relatively long.

4.     **PPC method applied to the electron trapping.**

While the photon induced perturbation of the geminate dynamics in systems like sc-$CO_2$ is difficult, the PPC method can still be used provided that (i) the conductivity signal is acquired in a time-resolved fashion and/or (ii) the density of the ionization events is sufficiently high so that some quasifree electrons decay by homogeneous recombination with holes in the solvent bulk. Given that these quasifree electrons are extremely mobile, this density does not have to be excessively high: e.g., in pulse radiolysis of sc-$CO_2$, such cross recombination was observed, albeit indirectly, when the electron concentration was in the micromolar range. [18] As for the time resolution, it does not have to be femto- or pico- seconds, as the only requirement to the conductivity setup is the possibility to distinguish between the prompt conductivity signal from $e_{qf}^-$ (that follows the convolution of the excitation pulse with the response function of the conductivity setup) and the long-lived conductivity signal from the anions. Specifically, the response time of the detection system should be several times shorter than $\mu_e\tau_e/\mu_s$. For sc-$CO_2$, this ratio is ca. 150 ns, and a time resolution of a few nanoseconds, which is quite standard in photoconductivity studies, [17,20] is more than adequate. This situation is illustrated in Fig. 1, where the "spike"-like prompt signal from quasifree electron generated by the 248 nm pulse (used for photoionization) and/or 532 nm pulse (used to detach the electron from the solvent anion) is readily time resolved against the background of the weaker, long-lived conductivity signal from the solvent anions. In the PPC experiments of Lukin and coworkers, [39] the typical load resistance of the conductivity cell was 1-2 GΩ and the time resolution was in the millisecond range. The high impedance made it possible to detect picomolar concentrations of the charge carriers; however, this high sensitivity was offset by inferior time resolution. However, for the measurement of the free electron yield, time resolution is not required.

Consider the simplest model in which a quasifree electron with the lifetime $\tau_e$ and mobility $\mu_e$ yields the solvent anion (or trapped electron) with mobility $\mu_s \ll \mu_e$. We will assume that the geminate stage is much shorter than $\tau_e$ (i.e., $\tau_e/t_c \gg 1$). Neglecting the cross recombination, the concentration $E$ of these electrons decreases with time as

$$E = E_0 \exp(-t/\tau_e) \tag{9}$$

while the concentration $A(t) = E_0 - E(t)$ of the solvent anions increases as

$$A(t) = E_0 \left[1 - \exp(-t/\tau_e)\right] \tag{10}$$

where $E_0$ is the initial concentration of the (free) electrons. The prompt conductivity signal $\sigma_e(t)$ from the electrons is given by

$$\sigma_e(t) = F\mu_e \, E(t) \tag{11}$$

where $F$ is the Faraday constant. The integral $S_p$ of this prompt signal is given by

$$S_p = \int_0^\infty \sigma_e(t) \, dt = F\mu_e\tau_e \, E_0 \tag{12}$$

Note, that the integral $S_p$ is not changed by the response function of the detection system. For $t \gg \tau_e$, the conductivity signal $\sigma_i$ from the solvent anions is given by

$$\sigma_i = F\mu_s \ A(t=\infty) = F\mu_s \ E_0, \tag{13}$$

so that the ratio

$$R = S_p/\sigma_i = \mu_e \tau_e / \mu_s. \tag{14}$$

This ratio also does not depend on the response function of the conductivity setup. The ratio $R$ can be determined more accurately than the quantities $S_p$ and $\sigma_i$ separately, because shot-to-shot variation in the electron concentration is divided out. The integral $S_p$ can be obtained by sampling the prompt conductivity signal using a boxcar integrator.

At a given delay time $t = \tau$ after the ionizing pump pulse the photoexcitation pulse with duration $\delta t \ll \tau_e$ converts a fraction $\phi$ of the solvent anions to the electrons:

$$E(\tau + \delta t) \approx E(\tau) + \phi A(\tau), \tag{15}$$

so that for $t > \tau$, the electron concentration $\tilde{E}(t)$ after the photoexcitation decays as

$$\tilde{E}(t) = [E(\tau) + \phi A(\tau)]\exp[-(t-\tau)/\tau_e]. \tag{16}$$

Since $E(t) = E(\tau)\exp[-(t-\tau)/\tau_e]$, the difference

$$\tilde{E}(t) - E(t) = \phi A(\tau)\exp[-(t-\tau)/\tau_e]. \tag{17}$$

Integrating both sides of eq. (17) from $\tau$ to infinity and substituting the result into eq. (12), we obtain that the integral of the prompt signal $S_p$ from the electrons increases by

$$\Delta S_p(\tau) = F \ \mu_e \tau_e \ \phi \ \beta \ A(\tau) \tag{18}$$

(where the coefficient $\beta = 1$) so that

$$\Delta S_p(\tau)/S_p = \phi \ \beta \ A(\tau)/A(t = \infty) \qquad (19)$$

Since the concentration of anions at $t = \infty$ is not changed after the photoexcitation, $\Delta \sigma_i(\tau) = 0$ and

$$\Delta R(\tau)/R = \phi \ A(\tau)/A(t = \infty) \qquad (20)$$

Thus, plotting the ratios $\Delta S_p(\tau)/S_p$ and $\Delta R(\tau)/R$ as a function of the delay time $\tau$ of the photoexcitation pulse yields the kinetics $A(\tau)$ of the anion formation (that is, the kinetics for electron trapping).

The obvious deficiency of this method is that improvement in the time resolution of the detection system can only be made by lowering the sensitivity. That, in turn, requires higher concentration of the photogenerated species. As discussed above, cross recombination of quasifree electrons might then become a concern. The surprising fact is that for $\phi << 1$, the same eqs. (19) and (20) are obtained even when this cross recombination is occurring in the photosystem (see the Appendix), with the coefficient $\beta$ in eq. (19) given by

$$\beta^{-1} = 1 + S_p/\varepsilon\varepsilon_0 = 1 + \left(\mu_e \tau_e / \mu_s\right) \ \sigma_i/\varepsilon\varepsilon_0. \qquad (21)$$

Since some quasifree electrons generated by the photoexcitation of solvent anions decay via homogeneous recombination, a *negative* net change $\Delta\sigma_i(\tau)$ in the conductivity signal $\sigma_i$ from anions is induced by the photoexcitation. As shown in the Appendix,

$$\Delta\sigma_i(\tau)/\sigma_i = -\phi \ (1-\beta) \ A(\tau)/A(t = \infty) \qquad (22)$$

i.e., the relative change in the long-lived conductivity signal follows the same formation kinetics $A(\tau)$ as the quantities given by eqs. (19) and (20). This long-lived conductivity signal can be sampled using high-impedance conductivity cells, in the same fashion as in the PPC experiments of Lukin and coworkers [39] and Braun and Scott. [38]

**5. Conclusion.**

It is shown that pump-probe photoconductivity method can be extended to study electron-trapping dynamics in nonpolar molecular liquids, provided that the time resolution of the conductivity setup is better than the ratio $\mu_e \tau_e / \mu_s$ of the mobility-lifetime product for the quasifree electron and the mobility $\mu_s$ of the trapped electron. For some liquids, this time is sufficiently long (e.g., ca. 150 ns in supercritical $CO_2$, see section 3) to use the standard low-impedance photoconductivity cells. In the situation when the time resolution cannot be increased (due to the adverse effect on the sensitivity), the electron trapping dynamics can still be studied provided that the trapping competes with the recombination of electrons and holes in the solvent bulk. Given that the mobility of quasifree electrons is very large (10-10$^3$ cm$^2$/Vs), this recombination is facile even when the density of the ionization events is relatively low. Perturbation of the geminate electron-hole dynamics is not needed for this PPC method.

## 6. Acknowledgement.

IAS thanks Dr. M. C. Sauer, Jr. for technical assistance. The research at the ANL was supported by the Office of Science, Division of Chemical Sciences, US-DOE under contract number W-31-109-ENG-38.

## 7. Appendix.

Let $E(t)$ be the concentration of free electrons, $A(t)$ be the concentration of the solvent anions, $\tau_e$ be the life time, and $k_2$ be the rate of bimolecular recombination of the quasifree electrons. Neglecting the recombination of the (relatively) low-mobility anions that occurs on a much longer time scale, we write

$$dE/dt = -E/\tau_e - k_2 \ E \ (E + A), \qquad (A1)$$

$$dA/dt = E/\tau_e, \qquad (A2)$$

where $E + A$ is the concentration of holes. Let us introduce dimensionless parameters $e = E/E_0$, $a = A/E_0$, $q = k_2 \tau_e E_0$, and $\eta = t/\tau_e$, where $E_0$ is the initial concentration of the electrons by the time $t \approx 0$ when the geminate recombination is complete. Eqs. (A1) and (A2) may now be rewritten as

$$de/d\eta = -\ e - q\ e\ (e + a), \quad (A3)$$

$$da/d\eta\ =\ e. \quad (A4)$$

Introducing the reduced concentration $c = e + a$ of the holes, we obtain, by summing up eqs. (A3) and (A4).

$$dc/d\eta = -\ q\ e\ c, \quad (A5)$$

or

$$d\ \ln c/d\eta\ =\ -q\ e. \quad (A6)$$

Integrating both parts of this equation and then using formula $a = \int e\ d\eta$ obtained by integration of both sides of eq. (A4), we have

$$c/c_i\ =\ \exp(-q[a - a_i]), \quad (A7)$$

where $a_i$ and $c_i$ are the reduced concentrations at some instant of time $t = t_i$. Since

$$c\ =\ a\ +\ da/d\eta \quad (A8)$$

and $da/d\eta = e = 0$ for $\eta \to \infty$, the final reduced concentration $a_\infty = a(t \to \infty)$ of the solvent anions is the root of equation

$$a_\infty\ =\ \exp(-q[a_\infty - a_i]). \quad (A9)$$

For $t_i = 0$, $a_i = 0$ and eq. (A9) simplifies to

$$a_\infty\ =\ \exp(-qa_\infty). \quad (A10)$$

Therefore, $a_\infty = \Phi(q)$, where the function $\Phi(q)$ is the root of eq. (A10) for a given $q$; for $q \ll 1$, $\Phi(q) \approx 1 - q$. The area $S_p$ under the prompt conductivity signal $\sigma_e(t)$ from the electrons is given by

$$S_p = F\mu_e \int_0^\infty E\, dt = F\left(\mu_e \tau_e\right) E_0\, \Phi(q), \tag{A11}$$

where we have used the equality $a_\infty = \int_0^\infty e\, d\eta$ obtained by integration of eq. (A4) from zero to infinity. The conductivity signal $\sigma_i$ from the (long-lived) ions at $t = \infty$ is given by

$$\sigma_i = F\mu_s\, E_0\, \Phi(q), \tag{A12}$$

so that the ratio $S_p/\sigma_i = \mu_e \tau_e / \mu_s$ does not depend on the recombination rate. Let us consider how the quantities $S_p$ and $\sigma_i$ change when a short laser pulse at $t = \tau$ promotes the electron from the solvent anion back into the conduction band. This pulse instantaneously converts a small fraction $\phi$ of the solvent anions to quasifree electrons,

$$a_\tau \xrightarrow{h\nu} a_\tau^\phi = a_\tau(1 - \phi) \quad and \quad e_\tau \xrightarrow{h\nu} e_\tau^\phi = e_\tau + a_\tau \phi, \tag{A13}$$

where $e_\tau$ and $a_\tau$ ($e_\tau^\phi$ and $a_\tau^\phi$) are the reduced electron and anion concentrations before and after this pulse, respectively. The photoconversion efficiency $\phi$ is the product of the photon fluence and the cross section for electron photodetachment. Since the total concentration $c_\tau = e_\tau + a_\tau$ of the charged species does not change after this photoexcitation, we once more use eq. (A9) to obtain

$$a_\infty^\phi = c_\tau\, \exp\left(-q\left[a_\infty^\phi - a_\tau^\phi\right]\right), \tag{A14}$$

where $a_\tau^\phi = a(t \to \infty)$ for the initial conditions given by eq. (A13). Substituting $\phi = 0$ in the latter equation, we obtain

$$a_\infty = c_\tau\, \exp\left(-q\left[a_\infty - a_\tau\right]\right). \tag{A15}$$

From eq. (A10) it follows that $c_\tau = \exp(-qa_\tau)$. Substituting the latter identity into eqs. (A13) and (A14) one obtains

$$a_\infty^\phi = \exp\left(-q\left[a_\infty^\phi - \phi\, a_\tau\right]\right). \tag{A16}$$

For $\phi \ll 1$, $a_\infty^\phi \approx a_\infty + \phi \left( \partial a_\infty^\phi / \partial \phi \right)_{\phi=0}$. The second term of this expansion can be determined by taking the derivatives of both parts of eq. (A16) at $\phi = 0$,

$$\left( \partial a_\infty^\phi / \partial \phi \right)_{\phi=0} = -q \; a_\infty \left\{ \left( \partial a_\infty^\phi / \partial \phi \right)_{\phi=0} + a_\tau \right\}, \tag{A17}$$

which gives

$$\left( \partial a_\infty^\phi / \partial \phi \right)_{\phi=0} = - \; G(q) \; a_\tau, \tag{A18}$$

where a new function

$$G(q) = q\Phi(q) / \left[ 1 + q\Phi(q) \right] \tag{A19}$$

is introduced. For $q \to 0$, $G(q) \to 0$. By integrating both parts of eq. (A4) from $t = \tau$ to infinity, the integral of $E(t)$ needed to estimate the quantity $S_p$ from eq. (A11) can be expressed as

$$\int_\tau^\infty E \; dt \; = \; E_0 \; \tau_e \left[ a_\infty^\phi - a_\tau^\phi \right]. \tag{A20}$$

From eq. (A13), we obtain the identity

$$\left( \partial a_\tau^\phi / \partial \phi \right)_{\phi=0} = -a_\tau. \tag{A21}$$

Substituting eqs. (A18) and (A21) into eq. (A20), we obtain

$$\left( \frac{\partial}{\partial \phi} \int_0^\infty E \; dt \right)_{\phi=0} = \left( \frac{\partial}{\partial \phi} \int_\tau^\infty E \; dt \right)_{\phi=0} = \frac{E_0 \tau_e a_\tau}{1 + q\Phi(q)}. \tag{A22}$$

Using eqs. (A22) and (A11), the photoinduced change $\Delta S_p(\tau) \approx \phi \left( \partial S_p / \partial \phi \right)_{\phi=0}$ in the integral $S_p$ induced by laser photoexcitation of the anion at the delay time $t = \tau$ is, therefore, given by

$$\Delta S_p(\tau) \approx F \left( \mu_e \tau_e \right) \; \phi \; \left[ 1 + q\Phi(q) \right]^{-1} \; A(\tau), \tag{A23}$$

where $A(\tau)$ is the anion concentration at $t = \tau$, while the photoinduced change

$$\Delta\sigma_i(\tau) \approx \phi(\partial\sigma_i/\partial\phi)_{\phi=0} = F\ \mu_s\ E_0\ \phi\ (\partial a_\infty^\phi/\partial\phi)_{\phi=0} \tag{A24}$$

in the signal $\sigma_i$ (eq. (A18)) is given by

$$\Delta\sigma_i(\tau) \approx -\ F\ \mu_s\ \phi\ G(q)\ A(\tau) \tag{A25}$$

Thus, *for $\phi \ll 1$, both $\Delta S_p(\tau)$ and $\Delta\sigma_i(\tau)$ are proportional to the instant anion concentration $A(\tau)$.* Numerical simulations indicate that eqs. (A23) and (A25) are accurate within 5% for $\phi < 0.5$. The proportionality coefficients in eqs. (A23) and (A25) depend on the dimensionless parameter $q = k_2 \tau_e E_0$. The product $q\Phi(q)$ can be estimated without knowing the quantities $E_0$, $\tau_e$, and $k_2$ separately. By the Debye equation, $k_2 = F\mu_e/\varepsilon\varepsilon_0$, where $\varepsilon$ is the dielectric constant and $\varepsilon_0$ is the permittivity of vacuum. Therefore,

$$q = F(\mu_e \tau_e) E_0/\varepsilon\varepsilon_0, \tag{A25}$$

and from eq. (A11) it follows that $q\Phi(q) = S_p/\varepsilon\varepsilon_0$. Substituting this identity into eqs. (A19), (A23) and (A25), eqs. (19) to (22) are obtained.

**Figure captions.**

**Figure 1.**

Laser-induced d.c. conductivity observed from 0.1 mM benzene in liquid-like sc-$CO_2$ ($\rho = 0.83$ g/cm$^3$, $T = 41^oC$); after ref. [17]. Benzene was photoionized using a 248 nm laser pulse (L2; 16 ns fwhm); the solvent anion was subsequently photoexcited using a 532 nm, 5 ns fwhm laser pulse (L2; 5 ns fwhm). Kinetics obtained for two delay times of the 532 nm pulse are shown. The "spikes" observed during the 248 nm and 532 nm photoexcitation are from quasifree electrons generated by biphotonic ionization of the solute end electron photodetachment from the solvent anion, respectively. The area of the "spike" in this plot gives the quantity $S_p$ given by eq. (12). The slow conductivity signal $\sigma_i$ is from solvent anions. The time profile of the "spike" follows the time profile of the excitation laser convoluted with the response function of the conductivity setup.

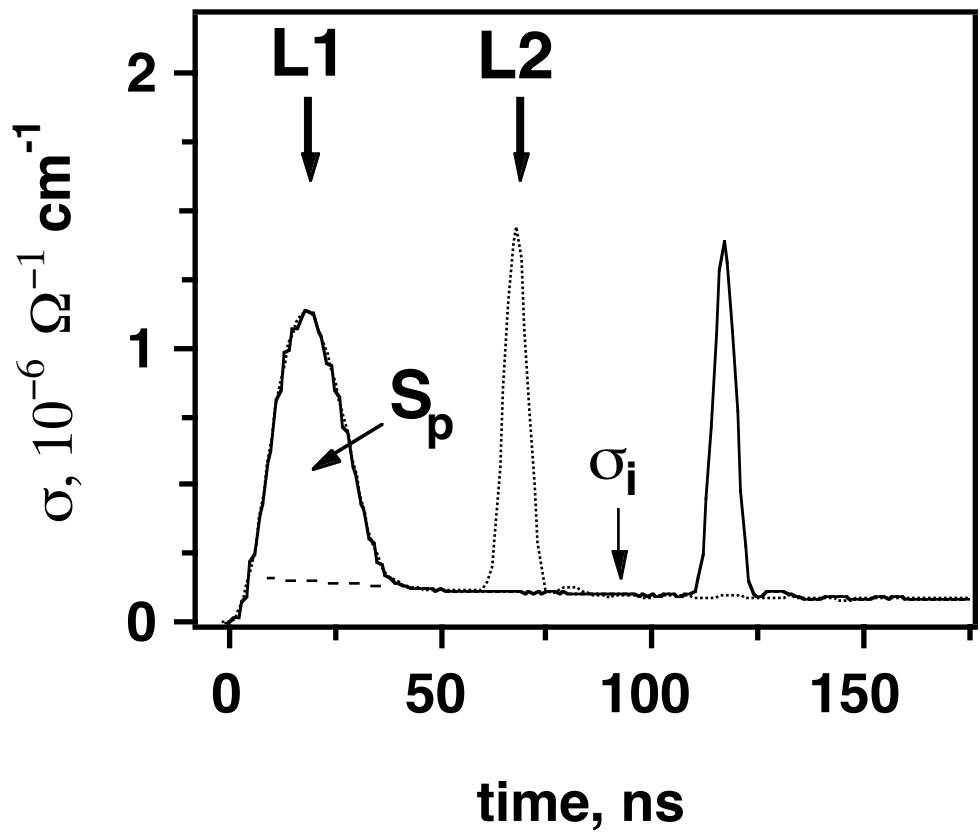

Figure 1. Shkrob & Ryzhik